\documentstyle[twoside,epsf,psfig,times]{article}
\def\beqa{\begin{eqnarray}}
\def\eeqa{\end{eqnarray}}
\def\beq{\begin{equation}}
\def\eeq{\end{equation}}
\catcode`\@=11
\long\def\@makefntext#1{
\protect\noindent \hbox to 3.2pt {\hskip-.9pt  
$^{{\eightrm\@thefnmark}}$\hfil}#1\hfill}               

\def\@makefnmark{\hbox to 0pt{$^{\@thefnmark}$\hss}}    

\def\ps@myheadings{\let\@mkboth\@gobbletwo
\def\@oddhead{\hbox{}
\rightmark\hfil\eightrm\thepage}   
\def\@oddfoot{}\def\@evenhead{\eightrm\thepage\hfil
\leftmark\hbox{}}\def\@evenfoot{}
\def\sectionmark##1{}\def\subsectionmark##1{}}

\oddsidemargin=\evensidemargin
\newcounter{sectionc}\newcounter{subsectionc}\newcounter{subsubsectionc}
\renewcommand{\section}[1] {\vspace{12pt}\addtocounter{sectionc}{1} 
\setcounter{subsectionc}{0}\setcounter{subsubsectionc}{0}\noindent 
        {\tenbf\thesectionc. #1}\par\vspace{5pt}}
\renewcommand{\subsection}[1] {\vspace{12pt}\addtocounter{subsectionc}{1} 
\setcounter{subsubsectionc}{0}\noindent 
{\bf\thesectionc.\thesubsectionc. {\kern1pt \bfit #1}}\par\vspace{5pt}}
\renewcommand{\subsubsection}[1] {\vspace{12pt}\addtocounter{subsubsectionc}{1}
        \noindent{\tenrm\thesectionc.\thesubsectionc.\thesubsubsectionc.
        {\kern1pt \tenit #1}}\par\vspace{5pt}}
\newcommand{\nonumsection}[1] {\vspace{12pt}\noindent{\tenbf #1}
        \par\vspace{5pt}}

\newcounter{appendixc}
\newcounter{subappendixc}[appendixc]
\newcounter{subsubappendixc}[subappendixc]
\renewcommand{\thesubappendixc}{\Alph{appendixc}.\arabic{subappendixc}}
\renewcommand{\thesubsubappendixc}
        {\Alph{appendixc}.\arabic{subappendixc}.\arabic{subsubappendixc}}

\renewcommand{\appendix}[1] {\vspace{12pt}
        \refstepcounter{appendixc}
        \setcounter{figure}{0}
        \setcounter{table}{0}
        \setcounter{lemma}{0}
        \setcounter{theorem}{0}
        \setcounter{corollary}{0}
        \setcounter{definition}{0}
        \setcounter{equation}{0}
        \renewcommand{\thefigure}{\Alph{appendixc}.\arabic{figure}}
        \renewcommand{\thetable}{\Alph{appendixc}.\arabic{table}}
        \renewcommand{\theappendixc}{\Alph{appendixc}}
        \renewcommand{\thelemma}{\Alph{appendixc}.\arabic{lemma}}
        \renewcommand{\thetheorem}{\Alph{appendixc}.\arabic{theorem}}
        \renewcommand{\thedefinition}{\Alph{appendixc}.\arabic{definition}}
        \renewcommand{\thecorollary}{\Alph{appendixc}.\arabic{corollary}}
        \renewcommand{\theequation}{\Alph{appendixc}.\arabic{equation}}
        \noindent{\tenbf Appendix \theappendixc #1}\par\vspace{5pt}}
\newcommand{\subappendix}[1] {\vspace{12pt}
        \refstepcounter{subappendixc}
        \noindent{\bf Appendix \thesubappendixc. {\kern1pt \bfit #1}}
        \par\vspace{5pt}}
\newcommand{\subsubappendix}[1] {\vspace{12pt}
        \refstepcounter{subsubappendixc}
        \noindent{\rm Appendix \thesubsubappendixc. {\kern1pt \tenit #1}}
        \par\vspace{5pt}}

\newcommand{\textlineskip}{\baselineskip=13pt}
\newcommand{\smalllineskip}{\baselineskip=10pt}

\newcommand{\copyrightheading}[1]
        {\vspace*{-2.5cm}\smalllineskip{\flushleft
        {\footnotesize International Journal of Modern Physics D, #1}\\
        {\footnotesize \copyright\kern2pt World Scientific Publishing
         Company}\\
         }}

\newcommand{\publisher}[2]{{\begin{center}\footnotesize\smalllineskip 
        Received #1\\
        Revised #2
        \end{center}
        }}

\def\abstracts#1#2#3{{
        \centering{\begin{minipage}{4.5in}\footnotesize\baselineskip=10pt
        \parindent=0pt #1\par 
        \parindent=15pt #2\par
        \parindent=15pt #3
        \end{minipage}}\par}} 

\def\keywords#1{{
        \centering{\begin{minipage}{4.5in}\footnotesize\baselineskip=10pt
        {\footnotesize\it Keywords}\/: #1
         \end{minipage}}\par}}

\renewenvironment{thebibliography}[1]
        {\frenchspacing
         \ninerm\baselineskip=11pt
         \begin{list}{\arabic{enumi}.}
        {\usecounter{enumi}\setlength{\parsep}{0pt}     
         \setlength{\leftmargin 12.7pt}{\rightmargin 0pt}
         \setlength{\itemsep}{0pt} \settowidth
        {\labelwidth}{#1.}\sloppy}}{\end{list}}

\newcounter{itemlistc}
\newcounter{romanlistc}
\newcounter{alphlistc}
\newcounter{arabiclistc}

\newcommand{\fcaption}[1]{
        \refstepcounter{figure}
        \setbox\@tempboxa = \hbox{\footnotesize Fig.~\thefigure. #1}
        \ifdim \wd\@tempboxa > 5in
           {\begin{center}
        \parbox{5in}{\footnotesize\smalllineskip Fig.~\thefigure. #1}
            \end{center}}
        \else
             {\begin{center}
             {\footnotesize Fig.~\thefigure. #1}
              \end{center}}
        \fi}

\newcommand{\tcaption}[1]{
        \refstepcounter{table}
        \setbox\@tempboxa = \hbox{\footnotesize Table~\thetable. #1}
        \ifdim \wd\@tempboxa > 5in
           {\begin{center}
        \parbox{5in}{\footnotesize\smalllineskip Table~\thetable. #1}
            \end{center}}
        \else
             {\begin{center}
             {\footnotesize Table~\thetable. #1}
              \end{center}}
        \fi}

\def\@citex[#1]#2{\if@filesw\immediate\write\@auxout
        {\string\citation{#2}}\fi
\def\@citea{}\@cite{\@for\@citeb:=#2\do
        {\@citea\def\@citea{,}\@ifundefined
        {b@\@citeb}{{\bf ?}\@warning
        {Citation `\@citeb' on page \thepage \space undefined}}
        {\csname b@\@citeb\endcsname}}}{#1}}

\newif\if@cghi
\def\cite{\@cghitrue\@ifnextchar [{\@tempswatrue
        \@citex}{\@tempswafalse\@citex[]}}
\def\citelow{\@cghifalse\@ifnextchar [{\@tempswatrue
        \@citex}{\@tempswafalse\@citex[]}}
\def\@cite#1#2{{$\null^{#1}$\if@tempswa\typeout
        {IJCGA warning: optional citation argument 
        ignored: `#2'} \fi}}

\def\pmb#1{\setbox0=\hbox{#1}
        \kern-.025em\copy0\kern-\wd0
        \kern.05em\copy0\kern-\wd0
        \kern-.025em\raise.0433em\box0}

\def\fnt#1#2{\footnotetext{\kern-.3em
        {$^{\mbox{\scriptsize #1}}$}{#2}}}

\def\fpage#1{\begingroup
\voffset=.3in
\thispagestyle{empty}\begin{table}[b]\centerline{\footnotesize #1}
        \end{table}\endgroup}

\def\runninghead#1#2{\pagestyle{myheadings}
\markboth{{\protect\footnotesize\it{\quad #1}}\hfill}
{\hfill{\protect\footnotesize\it{#2\quad}}}}
\font\tenrm=cmr10
\font\tenit=cmti10 
\font\tenbf=cmbx10
\font\bfit=cmbxti10 at 10pt
\font\ninerm=cmr9

\font\eightrm=cmr8

\def\qed{\hbox{${\vcenter{\vbox{                  
   \hrule height 0.4pt\hbox{\vrule width 0.4pt height 6pt
   \kern5pt\vrule width 0.4pt}\hrule height 0.4pt}}}$}}

\begin{document}
\setlength{\textheight}{7.7truein}    

\runninghead{G. S. Khadekar, A. Pradhan and M. R. Molaei}
{Higher Dimensional Dust Cosmological Implications of a Decay Law .... }
\normalsize\textlineskip
\thispagestyle{empty}
\setcounter{page}{1}

\copyrightheading{}             {Vol.~0, No.~0 (2005) 000--000}

\vspace*{0.88truein}

\fpage{1}

\centerline{\bf HIGHER DIMENSIONAL DUST COSMOLOGICAL IMPLICATIONS OF A DECAY LAW}
\vspace*{0.035truein}
\centerline{\bf FOR $\Lambda$ TERM : EXPRESSIONS FOR SOME OBSERVABLE QUANTITIES}
\vspace*{10pt}
\centerline{G. S. KHADEKAR}
\vspace*{0.015truein}
\centerline{\it Department of Mathematics, Nagpur University, Mahatma Jyotiba Phule 
Educational Campus,}
\baselineskip=10pt
\centerline{\it Amravati Road, Nagpur - 440 033, India}
\baselineskip=10pt
\centerline{\it gkhadekar@yahoo.com, gkhadekar@rediffmail.com}
\vspace*{10pt}
\centerline{ANIRUDH PRADHAN}
\vspace*{0.015truein}
\centerline{\it Department of Mathematics, Hindu Post-graduate College,}
\baselineskip=10pt
\centerline{\it Zamania, Ghazipur 232 331, India}
\baselineskip=10pt
\centerline{\it pradhan@iucaa.ernet.in, acpradhan@yahoo.com}
\vspace*{10pt}
\centerline{M. R. MOLAEI}
\vspace*{0.015truein}
\centerline{\it Department of Mathematics, University of Kerman, Kerman, Iran}
\baselineskip=10pt
\centerline{\it molaei\_mreza@yahoo.com}
\vspace*{0.225truein}
\publisher{(received date)}{(revised date)}
\vspace*{0.21truein}
\abstracts{In this paper we have considered the multidimensional cosmological implications of a 
decay law for  $\Lambda$ term that is proportional to $\beta \frac{\ddot {a}}{a}$, where $\beta$ 
is a constant and $a$ is the scale factor of RW-space time. We discuss the cosmological consequences 
of a model for the vanishing pressure for the case $k=0$. It has been observed that such models
are compatible with the result of recent observations and cosmological term $\Lambda$ gradually 
reduces as the universe expands. In this model $\Lambda$ varies as the inverse square of
time, which matches its natural units. The proper distance, the luminosity distance-redshift, 
the angular diameter distance-redshift, and look back time-redshift for the model are presented 
in the frame work of higher dimensional space time. The model of the Freese {\it et al.} 
({\it Nucl. Phys. B} {\bf 287}, 797 (1987)) for $n=2$ is retrieved for the particular choice of 
$A_{0}$ and also Einstein-de Sitter model is obtained for $A_{0} = \frac{2}{3}$. This work has thus 
generalized to higher dimensions the well-know result in four dimensional space time. It is found 
that there may be significant difference in principle at least, from the analogous situation in 
four dimensional space time. }{}{}
\vspace*{10pt}
\vspace*{10pt}

\vspace*{1pt}\textlineskip      
\section{Introduction}
\vspace*{-0.5pt}
\noindent
It is widely believed that a consistent unification of all fundamental forces in nature 
would be possible within the space-time with an extra dimensions beyond those four
observed so far. The absences of any signature of extra dimensions in current experiments 
is usually explained the compactness of extra dimensions. The idea of dimensional reduction 
or self compactification fits in particularly well in comology because if we belive in the 
big bang, our universe was much smaller  at the early stage and the present four dimensional 
stage could have been preceded by a higher dimensional one (Chodos and Detweiler\cite{ref1}). 
In this work we consider multidimensional Robertson Walker (RW) model as a test case. In RW 
type of homogenous cosmological model, the dimensionality has a marked effect on the time 
temperature relation of the universe and our universe appears to cool more slowly in higher 
dimensional space time (Chatterjee\cite{ref2}). 

In recent year, models with a relic cosmological constant $\Lambda$ have received considerable 
attention among researchers for various reasons (see Refs.\cite{ref3} $^-$ \cite{ref5} and 
references therein). We should realize that the existence of a nonzero cosmological constant in 
Einstein's equations is a feature of deep and profound consequence. The recent observations 
indicate that $\Lambda \sim 10^{-55}cm^{-2}$ while particle physics prediction for $\Lambda$ is 
greater than this value by a factor of order $10^{120}$. This discrepancy is known as cosmological 
constant problem. Some of the recent discussions on the cosmological constant ``problem'' an
consequence on cosmology with a time-varying cosmological constant are investigated by
Ratra and Peebles,\cite{ref6} Dolgov,\cite{ref7} $^-$ \cite{ref9} Sahni and Starobinsky,\cite{ref10} 
Padmanabhan\cite{ref11} and Peebles.\cite{ref12} For earlier reviews on this topic, the reader is 
referred to Zeldovich,\cite{ref13} Weinberg\cite{ref14} and Carroll, Press and Turner.\cite{ref15} 

Recent observations of type Ia supernovae (SNe Ia) at redshift $z < 1$  provide startling
and puzzling evidence that the expansion of the universe at the present time appears to be
{\it accelerating}, behaviour attributed to ``dark energy'' with negative pressure.
These observations (Perlmutter {\it et al.};\cite{ref16} Riess {\it et al.};\cite{ref17}
Garnavich {\it et al.};\cite{ref18} Schmidt {\it et al.}\cite{ref19}) strongly favour
a significant and positive value of $\Lambda$. The main conclusion of these observations is
that the expansion of the universe is accelerating.

A number of authors have argued in favour of the dependence $\Lambda \sim t^{-2}$ first 
expressed by Bertolami\cite{ref20} and later on by several authors\cite{ref21} $^-$ \cite{ref26} 
in different context. Recently, motivated by dimensional grounds in keeping with quantum cosmology, 
Chen and Wu,\cite{ref27} Abdel-Rahaman\cite{ref28} considered a $\Lambda$ varying as $a^{-2}$. 
Carvalho {\it et al.},\cite{ref29} Waga,\cite{ref30} Silveira and Waga,\cite{ref31} Vishwakarma\cite{ref32} 
have also considered/modified the same kind of variation. Such a dependence alleviates 
some problems in reconciling observational data with the inflationary universe scenario. Al-Rawaf and 
Taha and Al-Rawaf\cite{ref33} and Overdin and Cooperstock\cite{ref34} proposed a cosmological model 
with a cosmological constant of the form $\Lambda = \beta \frac{\ddot{a}}{a}$, where $\beta$
is a constant. Following the same decay law recently Arbab\cite{ref35} have investigated cosmic 
acceleration with positive cosmological constant and also analyze the implication of a model built-in 
cosmological constant for four-dimensional space time. The cosmological consequences of this decay law 
are very attractive. This law provides reasonable solutions to the cosmological puzzles presently
known. One of the motivations for introducing $\Lambda$ term is to reconcile the age parameter and the 
density parameter of the universe with recent observational data.

In this paper by considering cosmological implication of decay law for $\Lambda$ that proportional 
to $\frac{\ddot{a}}{a}$, we have calculated the deceleration parameter, age of the universe.   We
have also  analyzed the cosmological tests pertaining proper distance, luminosity distance, angular 
diameter distance, and look back time in the framework of higher dimensional space time for
dust model $p=0$ and shown that Freese {\it et al.}\cite{ref36} model is retrieved from our model 
for a particular choice of $A_{0}$ and $ n= 2$. The Einstein-de Sitter (ES) results are also
obtained from our results for the case $A_{0} = \frac{2}{3}$ and $n=2$.

\section{The Metric and Field  Equations}
Consider the $(n+2)$-dimensional homogeneous and isotropic model
of the universe represented by the space time
\begin{equation}
\label{eq1} ds^{2} = dt^2 - a^{2}(t)\left[\frac{dr^2}{1-k r^2}+
r^2 dX_{n}^{2}\right],
\end{equation}
where $a(t)$ is the scale factor, $k=0,\; \pm 1$ is the curvature
parameter and $$ dX_{n}^2= d\theta_{1}^2 + sin^2
\theta_{1}d\theta_{2}^2 +...+sin^2 \theta_{1}sin^2
\theta_{2}...sin^2 \theta_{n-1} d\theta_{n}^{2}$$.
The usual energy-momentum tensor is modified by addition of a term
\begin{equation}
\label{eq2}
T^{vac}_{ij} = - \Lambda(t) g_{ij},
\end{equation}
where $\Lambda(t)$ is the cosmological term and $g_{ij}$ is the
metric tensor. \\

Einstein's field equations (in gravitational units $c = 1, \; \; G = 1$) read as
\begin{equation}
\label{eq3}
 R_{ij} - \frac{1}{2} R g_{ij} = - 8 \pi
T_{ij}-\Lambda(t) g_{ij}.
\end{equation}
The energy-momentum tensor $T_{ij}$ in the presence of a perfect fluid has the form
\begin{equation}
\label{eq4}
T_{ij} = (p + \rho)u_{i}u_{j} - p g_{ij},
\end{equation}
where $p$ and $\rho$ are, respectively, the energy and
pressure of the cosmic fluid, and $u_{i}$ is the fluid
four-velocity such that $u^{i}u_{i} = 1$. \\
The Einstein filed Eqs. (3) and (4) for the metric (1)
take the form
\begin{equation}
\label{eq5}
\frac{n(n+1)}{2} \left[\frac{\dot{a}^{2}} {a^{2}}+
\frac{k}{a^2}\right]=8 \pi\rho + \Lambda(t),
\end{equation}
\begin{equation}
\label{eq6}
\frac{n\ddot{a}}{a} + \frac{n(n-1)}{2}\left[\frac{\dot{a}^{2}}
{a^{2}}+\frac{k}{a^2}\right] = - 8 \pi p + \Lambda(t).
\end{equation}
An over dot indicates a derivative with respect to time $t$. The
energy conservation equation $T^{i}_{j;i} = 0$ leads to
\begin{equation}
\label{eq7} \dot{\rho} + (n+1)(\rho + p)H = -\frac{\dot\Lambda}{8 \pi} \; ,
\end{equation}
where $H=\frac{\dot{a}}{a}$ is the Hubble parameter.\\

For complete determinacy of the system, we consider a perfect-gas
equation of state
\begin{equation}
\label{eq8} p = \gamma \rho, ~ ~ 0 \leq \gamma \leq 1.
\end{equation}
It is worth noting here that our approach suffers from a lack of
Lagrangian approach. There is no known way to present a consistent
Lagrangian model satisfying the necessary conditions discussed
in the paper.

\section{Solution of the Field Equations}
In case of the vanishing pressure i.e. $\gamma =0$ in Eq. (8),
Equations (6) with $k=0$ reduces to
\begin{equation}
\label{eq9} \frac{\ddot{a}}{a} + \frac{(n-1)}{2}\Big(\frac{\dot{a}}
{a}\Big)^{2} = \frac{\Lambda(t)}{n}.
\end{equation}
We propose a phenomenological decay law for $\Lambda$ of the form\cite{ref29,ref30}
\begin{equation}
\label{eq10} \Lambda = \beta \left(\frac{\ddot{a}}{a}\right),
\end{equation}
where $\beta$ is constant. Overdin and Cooperstock\cite{ref34} have pointed out that 
the model with $\Lambda \propto H^{2}$ is equivalent to above form. \\

Using Eq. (10) in Eq. (9) and by integrating we obtain
\begin{equation}
\label{eq11}
a(t) = \left[\frac{K}{A_{0}} ~ t \right]^{A_{0}},
\end{equation}
where $K$ is an integrating constant and the constant $A_{0}$ has the value
\begin{equation}
\label{eq12} A_{0} = \frac{2 (\beta -n)}{2\beta -n(n+1)}.
\end{equation}
By using Eq. (11) in the field equations Eqs. (5) and (6) we obtain
\begin{equation}
\label{eq13} \Lambda(t) =  \frac{2 n(n-1)\beta (\beta-n)}{[2\beta
-n(n+1)]^2} \frac{1}{t^{2}} \; ,\;\; \;\;\; \beta \ne
\frac{n(n+1)}{2}.
\end{equation}
\begin{equation}
\label{eq14}  \rho(t) =\frac{n(\beta-n)}{4\pi[2\beta -n(n+1)]}
\frac{1}{t^{2}} \; , \;\;\;\;\; \beta \ne \frac{n(n+1)}{2}.
\end{equation}
The deceleration parameter $q$ is defined as
\begin{equation}
\label{eq15} q = -\frac{\ddot{a}a}{\dot{a}^{2}}
=\frac{(1-A_{0})}{A_{0}}=\frac{n (n-1)}{2(n-\beta)} \;
,\;\;\;\;\;\beta \ne n
\end{equation}
The density parameter of the universe $\Omega_{m}$ is given by
\begin{equation}
\label{eq16} \Omega_{m} = \frac{16 \pi
\rho}{n(n+1)H^2}=\frac{2\beta -n(n+1)}{(n+1)(\beta-n)} \; ,
\;\;\;\;\;\beta \ne \frac{n}{2},  \; \; \; n \geq 2.
\end{equation}
The density parameter due to vacuum contribution is defined as
$\Omega_{\Lambda}=\frac{2\Lambda}{n(n+1)H^2}.$ Employing Eq. (13),
this gives
\begin{equation}
\label{eq17}
\Omega_{\Lambda}=\frac{(n-1)\beta}{(n+1)(\beta-n)}\;\;,\;\;\;\;\;\beta
\ne \frac{n}{2}, \; \; \; n \geq 2.
\end{equation}
From Eqs. (16) and (17), we obtain
 $$\Omega_{m} + \Omega_{\Lambda} = 1.$$
According to high redshift supernovae and CMB, the preliminary results from the advancing field 
of cosmology suggest that the universe may be accelerating universe with a dominant contribution 
to its energy density coming in the form of cosmological $\Lambda$-term. The results, when combined 
with CMB anisotropy observations on intermediate angular scales, strongly support a flat universe
\begin{equation}
\label{eq18} \Omega_{m} + \Omega_{\Lambda} = 1.
\end{equation}
The age of the universe is calculated as
\begin{equation}
\label{eq19} t_{0} = H_{0}^{-1}A_{0}.
\end{equation}
A high value of Hubble constant $H_{0} \sim $ 80 km/sec/Mpc  predicts a short age of the 
universe which is incompatible with the ages of oldest stars (12-16 Gyr) unless the universe 
is open $(\Omega_{m} < 0.1) $ or flat and $\Lambda$ is dominated  by $ \Omega_{m} +
\Omega_{\Lambda} = 1 $ (Sahni and Starobinsky\cite{ref10}).

\section{Neoclassical Tests (Proper Distance $d(z)$)}
A photon emitted by a source with coordinate $r=r_{1}$ and $t=t_{1}$ and received at a time $t_{0}$ 
by an observer located at $r=0$. The emitted radiation will follow null geodesics on which
$(\theta_{1},\theta_{2},...\theta_{n})$ are constant. 

The proper distance between the source and observer is given by
\begin{equation}
\label{eq20}
 d(z)=a_{0}\int_{a}^{a_{0}}\frac{da}{a \dot{a}}\;\;\:,
\end{equation}
$$r_{1}=\int_{t_{1}}^{t_{0}}\frac{dt}{a}=\frac{a_{0}^{-1}H_{0}^{-1}A_{0}}{(1-A_{0})}
\left[ 1- (1+z)^{\frac{A_{0}-1}{A_{0}}}\right].$$ Hence
\begin{equation}
\label{eq21}
 d(z)=r_{1}a_{0}=H_{0}^{-1}\left(\frac{A_{0}}{1-A_{0}}\right)
 \left[ 1- (1+z)^{\frac{A_{0}-1}{A_{0}}}\right],
\end{equation}
where $(1 + z) = \frac{a_{0}}{a}$ = redshift and $a_{0}$ is the present scale factor of the
universe.

For small $z$ Eq. (21) reduces to
 \begin{equation}
 \label{eq22}
 H_{0} d(z)= z - \frac{1}{2}A_{0}z^{2} + ...
\end{equation}
By using Eq. (15)
\begin{equation}
\label{eq23}
 H_{0} d(z)= z - \frac{1}{2}(1 + q)z^{2} + ...
\end{equation}
From Eq. (21), it is observed that the distance $d$ is maximum at
$z = \infty$. Hence
\begin{equation}
\label{eq24}
 d(z = \infty) = H_{0}^{-1}\left(\frac{A_{0}}{1-A_{0}}\right)=\frac{2H_{0}^{-1}(n-\beta)}{n(n-1)}
 \end{equation}
Eq. (21) gives the Freese {\it et al.} results for the proper
distance if we choose $n=2$ and
$$\frac{A_{0}}{(1 - A_{0})} =\frac{1}{q}=  \frac{2}{(3 \Omega_{m} - 2)}$$
and $d(z)$ is maximum for $\Omega_{m} \to 0 $ (de-Sitter
universe) and minimum for $\Omega_{m} \to 1 $ ES.

\section{Luminosity Distance}
Luminosity distance is the another important concept of
theoretical cosmology of a light source. The luminosity distance is a way of expanding
the amount of light received from a distant object. It is the distance that the object
appears to have, assuming the inverse square law for the reduction of light intensity
with distance holds. The luminosity distance is {\it not} the actual distance to the
object, because in the real universe the inverse square law does not hold. It is broken
both because the geometry of the universe need not be flat, and because the universe is
expanding. In other words, it is defined in such a way as generalizes the inverse-square
law of the brightness in the static Euclidean space to an expanding curved space (Waga\cite{ref30}).

If $d_{L}$ is the luminosity distance to the object, then
\begin{equation}
\label{eq25}
d_{L} = \left(\frac{L}{4\pi l}\right)^{\frac{1}{2}},
\end{equation}
where $L$ is the total energy emitted by the source per unit time,
$l$ is the apparent luminosity of the object. Therefore one can
write
\begin{equation}
\label{eq26}
d_{L} = r_{1}a_{0}= d(1 + z).
\end{equation}
Using Eq. (21) equation  (26) reduces to
\begin{equation}
\label{eq27}
 H_{0} d_{L} = (1 + z)\left(\frac{A_{0}}{1-A_{0}}\right)\Big[1 - (1 +
z)^{\frac{A_{0}-1}{A_{0}}}\Big].
\end{equation}
For small $z$, Eq. (27) gives
\begin{equation}
\label{eq28}
H_{0} d_{L} = z + \frac{1}{2}(1 - q)z^{2} + ...
\end{equation}
or by using Eq. (16)
\begin{equation}
\label{eq29}
H_{0} d_{L} = z + \Big[1-\left(\frac{n+1}{2}\right)\Omega_{m}\Big]z^{2} +...
\end{equation}
The luminosity distance depends on the cosmological model we have under discussion, and hence
can be used to tell us which cosmological model describe our universe. Unfortunately, however,
the observable quantity is the radiation flux density received from an object, and this can
only be translated into a luminosity distance if the absolute luminosity of the object is known.
There is no distant astronomical objects for which this is the cases. This problem can however be
circumvented if there are a population of objects at different distances which are believed to have
the same luminosity; even if that luminosity is not known, it will appear merely as an overall
scaling factor. 

Such a population object is Type Ia supernovae. These are believed to be caused by the core
collapse of white dwarf stars when they accrete material to take them over the Chandrasekhar
limit. Accordingly, the progenitor of such supernovae are expected to very similar, leading
to supernovae of a characteristic brightness. This already gives good standard candle, but
it can be further improved as there is an observed correlation between the maximum absolute
brightness of a supernova and the rate at which its brightens and faded. And because a supernova at
maximum brightness has a luminosity comparable to an entire galaxy, they can be seen at great
distance. Exactly such an effect has been observed for several dozen Type Ia high z supernovae
$(z_{max} \leq 0.83)$ by two teams: the supernova cosmology project\cite{ref16} and the high-z
supernova search team. \cite{ref17} The results delivered a major surprise to cosmologists.
None of the usual cosmological models without a cosmological constant were able to explain the
observed luminosity distance curve. The observations of Perlmutter {\it et al.}\cite{ref16} indicate
that the joint probability distribution of $\left(\Omega_{m}, \Omega_{\Lambda}\right)$ is well fitted by
$$ 0.8\Omega_{m}  - 0.6\Omega_{\Lambda} \simeq - 0.2 \pm 0.1.$$
The best-fit region strongly favours a positive energy density for the cosmological constant
$\Omega_{\Lambda} > 0$. 

\section{Angular Diameter Distance}
The angular diameter distance is a measure of how large objects appear to be. As with
the luminosity distance, it is defined as the distance that an object of known physical
extent appears to be at, under the assumption of Euclidean geometry. 

The angular diameter $d_{A}$ of a light source of proper distance
$d$ is given by
\begin{equation}
\label{eq30}
d_{A} = d(z)(1+z)^{-1}=d_{L}(1 + z)^{-2}.
\end{equation}
Applying Eq. (27) we obtain
\begin{equation}
\label{eq31}
H_{0} d_{L} =\frac{A_{0}}{1-A_{0}}\left[\frac{1 - (1
+ z)^{\frac{A_{0}-1}{A_{0}}}}{(1+z)}\right].
\end{equation}
Usually $d_{A}$ has a minimum (or maximum) for some $Z=
Z_{m}$. In Freese {\it et al.} \cite{ref36} model, for
example this occurs for $n=2$ and
$$Z_{m} =  \Big(\frac{3}{2}\Omega_{m}\Big)^{\frac{2}{(3\Omega_{m} - 2)}} - 1 .$$
The maximum $d_{A}$ for our model, one can easily find by setting the value of $\Omega_{m}$. 

The angular diameter and luminosity distances have similar forms, but have a different
dependence on redshift. As with the luminosity distance, for nearly objects the angular
diameter distance closely matches the physical distance, so that objects appear smaller
as they are put further away. However the angular diameter distance has a much more
striking behaviour for distant objects. The luminosity distance effect dims the radiation
and the angular diameter distance effect means the light is spread over a large angular area.
This is so-called surface brightness dimming is therefore a particularly strong function
of redshift. 

\section{Look Back Time}
The time in the past at which the light we now receive from a distant object was emitted is 
called the look back time. How {\it long ago} the light was emitted (the look back time)
depends on the dynamics of the universe. 

The radiation travel time (or look back time) $(t - t_{0})$ for photon emitted by a source at 
instant $t$ and received at $t_{0}$ is given by
\begin{equation}
\label{eq32} t - t_{0} = \int^{a_{0}}_{a} \frac{da}{\dot{a}} \; ,
\end{equation}
Equation (11) can be rewritten as
\begin{equation}
\label{eq33} a = B_{0} \: t^{A_{0}}, \;\; B_{0} = constant.
\end{equation}
This follows that
\begin{equation}
\label{eq34} \frac{a_{0}}{a} = 1 + z =
\left(\frac{t_{0}}{t}\right)^{A_{0}},
\end{equation}
The above equation gives
\begin{equation}
\label{eq35} t = t_{0}(1 + z)^{-\frac{1}{A_{0}}}.
\end{equation}
From Eqs. (21) and (35), we obtain
\begin{equation}
\label{eq36} t_{0} - t = A_{0}H_{0}^{-1}\left[1 - (1 +
z)^{-\frac{1}{A_{0}}}\right],
\end{equation}
which is
\begin{equation}
\label{eq37} H_{0} (t_{0} - t) = A_{0}\left[1 - (1 +
z)^{-\frac{1}{A_{0}}}\right].
\end{equation}
For small $z$ one obtain
\begin{equation}
\label{38} H_{0} (t_{0} - t) = z - (1+ \frac{q}{2}) z^{2} + ....
\end{equation}
From Eqs. (35) and (37), we observe that at $z \to \infty$, $H_{0}
t_{0}$ = $A_{0}$ (constant). For $n=2$ and  $A_{0}$ =
$\frac{2}{3}$ gives the well-known ES result
\begin{equation}
\label{eq39} H_{0} (t_{0} - t) =  \frac{2}{3}\left[1 - (1 + z)^{-\frac{3}{2}}\right].
\end{equation}
\section {Discussion}
In the above we have presented the cosmological consequences of Eq. (10) for the vanishing 
pressure in the frame work of higher dimensional space time. The zero pressure is not strictly 
appropriate because it occurs at the latter stage of evolution when ``extra'' dimension
lose much of their significance and the dimensionality has a marked effect on the time temperature 
relation of the universe, and our universe appears to cool more slowly in higher dimensional space 
time suggested by Chaterjee.\cite{ref2} To solve the age parameter and density parameter one require 
the cosmological constant to be positive or equivalently the deceleration parameter to be
negative. The nature of the cosmological constant
$\Lambda$ and the energy density $\rho$ have been examined. We have found that the cosmological 
parameter $\Lambda$ varies inversely with the square of time, which matches its natural units. 
This supports the views in favour of the dependence $\Lambda \sim t^{-2}$ first expressed by 
Bertolami\cite{ref20} and later on observed by several authors.\cite{ref21} $^-$ \cite{ref26}

We also attempted to investigate the well known astrophysical phenomena, namely the neoclassical 
test, the luminosity distance-redshift, the angular diameter distance-redshift, and look back 
time-redshift for the model in the frame work of multidimensional space time. The importance of 
the concept of luminosity distance to the study of theoretical astrophysics needs no further 
elaboration as the intrinsic luminosity of a source may be calculated with help of source's redshift 
and apparent luminosity are know. However,  as it is related to the scale factor, space curvature, etc. 
The results for the cosmological tests are compatible with the present observations. The model of
the Freese {\it et al.} is retrieved from our model for $n=2$  and the particular choice of $A_{0}$. 
Also the ES results are obtained for the case $A_{0} = \frac{2}{3}$. These tests are found to
depend on $\beta$. It is a general belief among cosmologists that more precise observational data should 
be achieved in order to make more definite statements about the validity of cosmological
models (Charlton and Turner\cite{ref37}). We hope that in the near future, with the new generation of 
the telescope, the present situation could be reversed. 

\nonumsection{Acknowledgements}
The authors (GSK and AP) wish to thank the Harish-Chandra Research Institute (HRI), Allahabad, India, 
for providing warm hospitality and excellent facilities where part of this work was done.
\newline
\newline
\nonumsection{References}

\end{document}